\documentclass[%
 aip,
 jmp,%
 amsmath,amssymb,
 reprint,%
]{revtex4-1}

\usepackage{listings}

\usepackage{subfigure}
\usepackage{graphicx}
\graphicspath{{./images_all/spin_circuit_STFMR/}}

\usepackage{dcolumn}
\usepackage{bm}

\begin{document}

\title{Spin-circuit representation of spin-torque ferromagnetic resonance}

\author{Kuntal Roy}
\email{kuntal@iiserb.ac.in}
\noaffiliation
\affiliation{Department of Electrical Engineering and Computer Science\\
Indian Institute of Science Education and Research Bhopal, Bhopal, Madhya Pradesh 462066, India}


\begin{abstract}
Spin-torque ferromagnetic resonance (ST-FMR) particularly using magnetic insulators and heavy metals possessing a giant spin Hall effect (SHE) has gotten a lot of attention for the development of spintronic devices. To devise complex functional devices, it is necessary to construct the equivalent spin-circuit representations of different phenomena. Such representation is useful to translate physical equations into circuit elements, benchmarking experiments, and then proposing creative and efficient designs. We utilize the superposition principle in circuit theory to separate the spin Hall magnetoresistance and spin pumping contributions in the ST-FMR experiments. We show that the proposed spin-circuit representation reproduces the standard results in literature. We further consider multilayers like a spin-valve structure with an SHE layer sandwiched by two magnetic layers and show how the corresponding spin-circuit representation can be constructed by simply writing a vector netlist and solved using circuit theory.
\end{abstract}



\maketitle

In ferromagnetic resonance (FMR) experiments,~\cite{RefWorks:1311,RefWorks:1308,RefWorks:1168,RefWorks:2636} a magnetization precesses around an effective magnetic field and the rotation is sustained by a transverse ac field acting as negative damping.~\cite{RefWorks:162,RefWorks:161} In spin-torque ferromagnetic resonance (ST-FMR) experiments, an in-plane alternating current in spin-orbit materials drives the magnetization precession in an adjacent magnet via direct spin Hall effect (SHE)~\cite{RefWorks:760,*RefWorks:764,*RefWorks:1198,RefWorks:771,*RefWorks:765,*RefWorks:902,RefWorks:769,roy14_3} and it pumps spins into the spin-orbit layer detected as a charge voltage/current via inverse spin Hall effect (ISHE).~\cite{RefWorks:1325,RefWorks:1111,RefWorks:1101,RefWorks:1278} Spin pumping~\cite{RefWorks:881,*RefWorks:1041,RefWorks:876} is the reciprocal phenomenon~\cite{RefWorks:1295} of spin-torque~\cite{RefWorks:8,*RefWorks:155} according to Onsager's reciprocity.~\cite{RefWorks:1292,*RefWorks:1293} Although ST-FMR was developed first for metallic magnets,~\cite{RefWorks:817,RefWorks:1333,RefWorks:1016} later it has been shown that ST-FMR is also possible for magnetic insulators e.g., yttrium-iron-garnet (YIG),~\cite{RefWorks:2591} which is promising for the development of spintronic devices. Spin Hall magnetoresistance (SMR)~\cite{RefWorks:1024,RefWorks:1019,RefWorks:1025,RefWorks:1026,RefWorks:1027,RefWorks:1028,RefWorks:1029} i.e., the dependence of electrical resistance of spin-orbit material on the magnetization direction has been observed for magnetic insulators due to the simultaneous action of SHE and ISHE, and it is modulated by the magnetization direction via the interfacial spin transfer. The fundamental difference of SMR with other magnetoresistances (anisotropic magnetoresistance (AMR),~\cite{amr} giant magnetoresistance (GMR),~\cite{RefWorks:434,*RefWorks:433} tunneling magnetoresistance (TMR)~\cite{RefWorks:577,*RefWorks:76,*RefWorks:33,*RefWorks:74}) is that current does not need to pass through the magnet. Both longitudinal SMR and spin pumping contribute to the dc voltage/current detected in ST-FMR experiments. 

Here we develop the spin-circuit representation of ST-FMR comprising both SMR and spin pumping. Circuit theory (Kirchhoff's current and voltage laws, KCL and KVL, originating from the conservation of charge and energy, respectively) has been tremendously successful for the development of the transistor-based technology.~\cite{rabae03} The physical equations are translated first into circuit elements for simplified understandings and it helps us to apply the principles of circuit theory e.g. superposition principle. Then we further analyze the circuit in this organized form, solve it programmatically, benchmark experiments, and can develop complex creative designs for applications.~\cite{rabae03} Due to the advent of new materials and phenomena in the field of spintronics, it needs to connect different spin-circuit representations of different phenomena and this is necessary for the commercial developments e.g., SPICE (Simulation Program with Integrated Circuit Emphasis).~\cite{hspice} Such spintronic circuit models apply to quantum transport~\cite{RefWorks:1350} as well. The voltages and currents at different nodes are of 4-components (1 for charge and 3 for spin vector) and the conductances are $4\times 4$ matrices ($c$-$z$-$x$-$y$ basis). Some spin-circuit representations that have been developed earlier are for ferromagnet (FM), normal metal (NM), FM-NM interface, spin Hall effect, and spin pumping.~\cite{RefWorks:198,srini14,RefWorks:1253,roy17_2}

\begin{figure*}
\centering
\includegraphics[width=\textwidth]{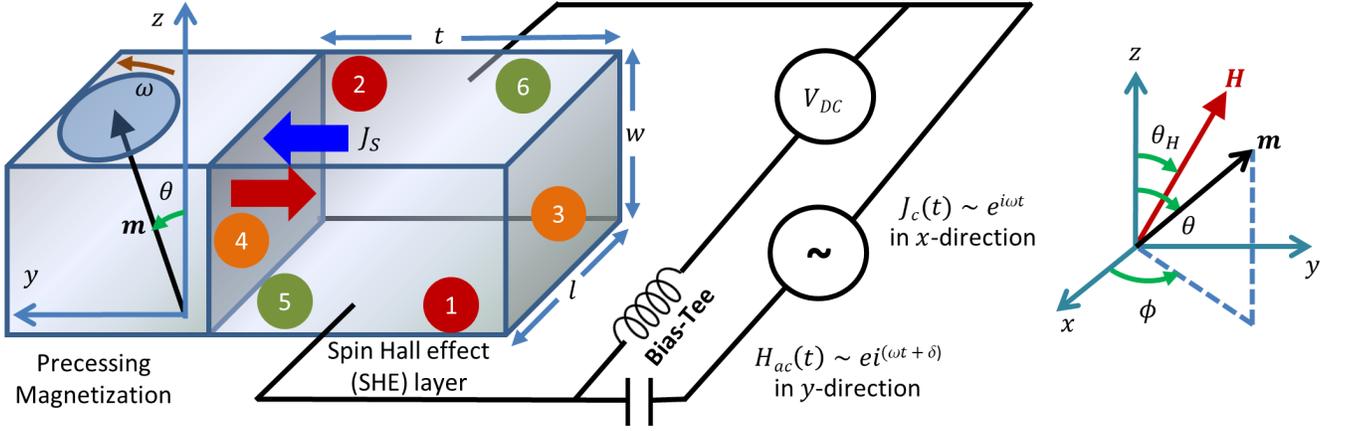}
\caption{\label{fig:ST-FMR} Schematic diagram of ST-FMR setup and axis assignments. With the help of a bias-tee, simultaneous ac transmission and dc voltage/current detection is possible in ST-FMR. A longitudinal ac current in $x$-direction $J_c$ across the surfaces marked by 1 and 2 of the SHE layer generates a transverse spin current density $J_s$ and it can drive the magnetization $\mathbf{m}$ into precession. $H$ is the applied dc magnetic field, $H_{ac}$ is the rf driving field produced from $J_c$ with a phase lag $\delta$, $\omega$ and $\theta$ are the precession frequency and angle, respectively. Spin potentials are developed across the surfaces marked by 3 and 4. Using a bias-tee, the dc voltage $V_{dc}$ can be detected and it comprises of both the contributions of longitudinal SMR and spin pumping due to ISHE. Also, charge potentials are developed across the surfaces marked by 5 and 6 due to transverse SMR. In the axis assignments, standard spherical coordinate system is used with azimuthal angle $\phi$ and polar angle $\theta$ for magnetization $\mathbf{m}$, and polar angle $\theta_H$ for the dc field $H$.}
\end{figure*}

Figure~\ref{fig:ST-FMR} shows the schematic diagram of the ST-FMR setup. A longitudinal ac current in the SHE layer having a length $l$, width $w$, and thickness $t$ in the presence of a dc magnetic field drives the magnetization in the adjacent magnetic layer into precession via SHE and utilizing a bias-tee we can detect the dc voltage/current in the longitudinal direction due to SMR and spin pumping. 

To develop the spin-circuit representation we can utilize the Ohm's law with spin-orbit interaction (including both SHE and ISHE)~\cite{RefWorks:1024}
\begin{equation}
\left\lbrack{\begin{array}{c}
  \mathbf{J^c} \\
  \mathbf{J^z} \\
	\mathbf{J^x} \\
	\mathbf{J^y} \\
  \end{array} } \right\rbrack
= \sigma
\left\lbrack {\begin{array}{cccc}
   1 & \theta_{sh}\mathbf{\hat{z}}\times & \theta_{sh}\mathbf{\hat{x}}\times & \theta_{sh}\mathbf{\hat{y}}\times \\
   \theta_{sh}\mathbf{\hat{z}}\times & 1 & 0 & 0\\
	 \theta_{sh}\mathbf{\hat{x}}\times & 0 & 1 & 0\\
	 \theta_{sh}\mathbf{\hat{y}}\times & 0 & 0 & 1\\
	 \end{array} } \right\rbrack
\left\lbrack{\begin{array}{c}
  -\mathbf{\nabla}V^c \\
  -\mathbf{\nabla}V^z \\
	-\mathbf{\nabla}V^x \\
	-\mathbf{\nabla}V^y \\
  \end{array} } \right\rbrack
\nonumber
\label{eq:Ohm_SOC}
\end{equation}
together with the charge and spin diffusion equations $\mathbf{\nabla \cdot J^c} = -\sigma\nabla^2 V^c = 0$ and $\mathbf{\nabla \cdot J^{x,y,z}} = -\sigma\nabla^2 V^{x,y,z} = - (\sigma/\lambda^2) V^{x,y,z}$, respectively, where $\mathbf{J^c}$ and $\mathbf{J^{x,y,z}}$ are current densities, and $V^c$ and $V^{x,y,z}$ are spin potentials, $\sigma$ is the conductivity, $\lambda$ is the spin diffusion length, and $\theta_{sh}$ is the spin Hall angle of the SHE layer.

Since we are not considering any charge component in the magnetic insulators, the generation of inverse spin Hall voltage is in the transverse direction due to the symmetry involved in the system, and unlike charge pumping,~\cite{RefWorks:1034,*RefWorks:1324} a precessing magnet injects a \emph{pure} spin current into surrounding conductors,~\cite{RefWorks:881,*RefWorks:1041,RefWorks:876} we consider a reduced 3-component version of the spin-circuit and show that it can reproduce the established expressions in literature for the dc voltage/current generated in ST-FMR experiments comprising both SMR and spin pumping. We further employ such formalism for multilayers like a spin-valve structure with an SHE layer sandwiched by two magnetic layers.~\cite{RefWorks:1024} We show how we can simply write a vector netlist comprising the conductances and voltage/current sources so that we can solve for the voltages/currents at the different nodes \emph{without} invoking any boundary condition using a circuit solver. For complex structures, derivation of any analytical expression becomes tedious and this signifies the prowess of the spin-circuit approach.

\begin{figure*}
\centering
\includegraphics[width=\textwidth]{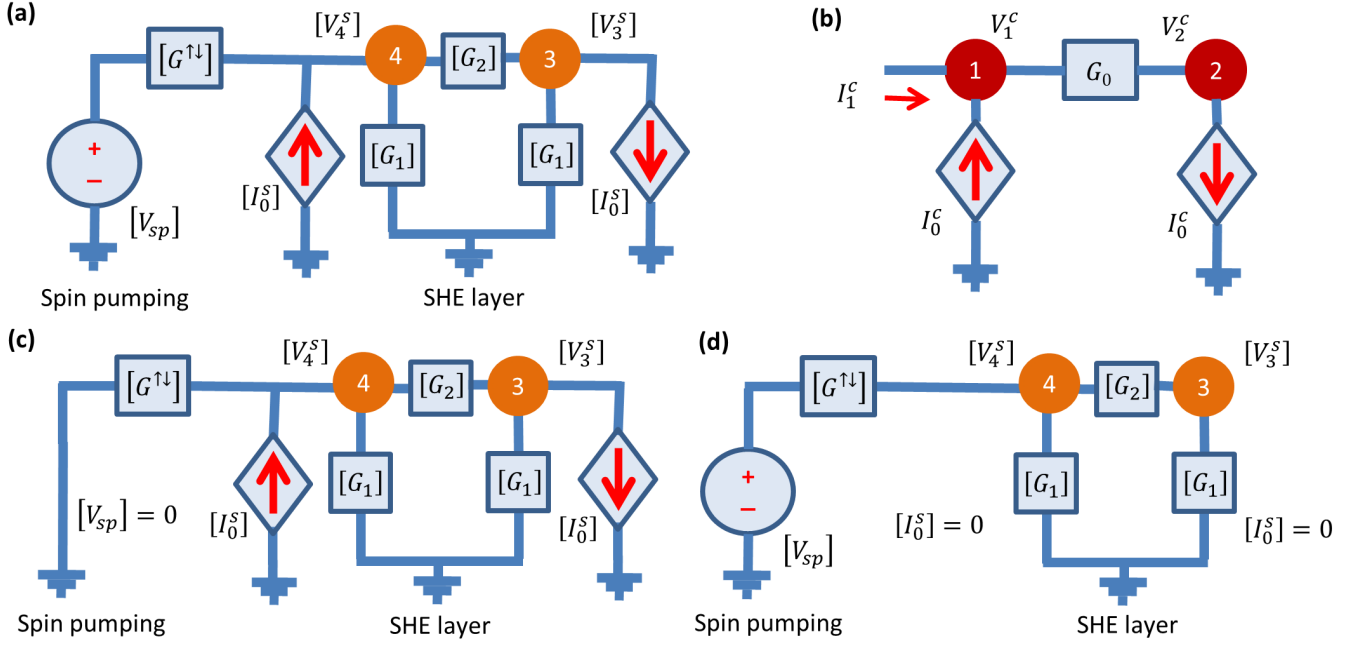}
\caption{\label{fig:ST-FMR_circuit} (a) Spin-circuit representation of ST-FMR. Spin potentials are developed at the surfaces marked by 3 and 4. $\left\lbrack I_{0}^s\right\rbrack$ is the charge current dependent spin current source due to SHE and $\left\lbrack V_{sp}\right\rbrack$ acts as the spin battery due to spin pumping. The bulk diffusion in the SHE layer can be represented by a $\pi$-circuit defined in the text. (b) The charge circuit for the spin-to-charge conversion by ISHE with the spin current dependent charge current sources $I_0^c$ with $G_0$ representing the conductance of the SHE layer. (c) Spin-circuit representation of ST-FMR without considering the spin battery $\left\lbrack V_{sp}\right\rbrack$. (d) Spin-circuit representation of ST-FMR without considering the spin current sources $\left\lbrack I_{0}^s\right\rbrack$.}
\end{figure*}

Figure~\ref{fig:ST-FMR_circuit}(a) shows the spin-circuit representation of ST-FMR that constitutes the spin-circuit representations of SHE layer, the charge current dependent spin current sources due to the charge to spin conversion in the SHE layer, and spin pumping. The instantaneous spin pumping can be represented by $\left\lbrack I_{sp}\right\rbrack=\left\lbrack G^{\uparrow\downarrow}\right\rbrack \left\lbrack V_{sp}\right\rbrack$, where $\left\lbrack V_{sp}\right\rbrack = (\hbar/2e) \, [0\;0\;1]^T$, $[G^{\uparrow\downarrow}]$ is the interfacial \emph{bare} complex spin mixing conductance between the magnetic layer and the SHE layer represented in $(\mathbf{m},\mathbf{m'},\mathbf{m} \times \mathbf{m'})$ basis (where $'$ denotes the time derivative) as
\begin{equation}
\left\lbrack G^{\uparrow\downarrow}\right\rbrack
=\left\lbrack\begin{array}{crr}
  0 & 0 & 0\\
	0 & 2G_{r} & 2G_{i}\\ 
  0 & -2G_{i} & 2G_{r}
\end{array} \right\rbrack,
\label{eq:G_SP}
\end{equation}
$\left\lbrack I_{sp}\right\rbrack = (\hbar/2e)\, [0\;2G_{i}\;2G_{r}]^T$ in $(\mathbf{m},\mathbf{m'},\mathbf{m} \times \mathbf{m'})$ basis, and $2G_{r(i)}$ is the real (imaginary) part of the \emph{bare} complex spin mixing conductance $G^{\uparrow\downarrow}$. The \emph{bulk} diffusion in the SHE layer can be represented by a $\pi$-circuit as shown in the Fig.~\ref{fig:ST-FMR_circuit}(a) with $\left\lbrack G_{1(2)} \right\rbrack=G_{1(2)} \left\lbrack I_{3\times3} \right\rbrack$, where $G_{1}=G_\lambda \mathrm{tanh} (t/2\lambda)$, $G_{2}=G_\lambda \mathrm{csch} (t/\lambda)$, $G_\lambda = \sigma l w/\lambda$, and $\left\lbrack I_{3\times3}\right\rbrack$ is the ${3\times3}$ identity matrix. The spin current sources in the Fig.~\ref{fig:ST-FMR_circuit}(a) in $z$-$x$-$y$ basis can be written as 
\begin{equation}
\left\lbrack I_{0,zxy}^s\right\rbrack = \beta G_0 \left(V_1^c - V_2^c \right) \left\lbrack 1 \; 0 \; 0\right\rbrack^T
\label{eq:I_0_zxy}
\end{equation}
where $G_0 = \sigma t w/l$ and $\beta = \theta_{sh} l/t$. After converting to $(\mathbf{m},\mathbf{m'},\mathbf{m} \times \mathbf{m'})$ basis, the Equation~\eqref{eq:I_0_zxy} becomes (see supplementary material)
\begin{equation}
\left\lbrack I_{0}^s\right\rbrack = I_{0}^{s0} m_z \mathbf{m} + I_{0}^{s0}\frac{1}{\omega}\mathbf{m} \times \mathbf{m'}
\end{equation}
where $I_{0}^{s0} = \beta G_0 \left(V_1^c - V_2^c \right)$.

Applying KCL at the nodes 3 and 4 in the spin-circuit shown in the Fig.~\ref{fig:ST-FMR_circuit}(a), we get (see supplementary material)
\begin{equation}
 \left\lbrack {\begin{array}{cc}
   \left\lbrack G_1\right\rbrack + \left\lbrack G_2\right\rbrack & -\left\lbrack G_2\right\rbrack \\
   -\left\lbrack G_2\right\rbrack & \left\lbrack G_1\right\rbrack + \left\lbrack G_2\right\rbrack + \left\lbrack G^{\uparrow\downarrow}\right\rbrack
 \\
  \end{array} } \right\rbrack
\left\lbrack{\begin{array}{c}
  \left\lbrack V_3^s\right\rbrack \\
  \left\lbrack V_4^s\right\rbrack \\
  \end{array} } \right\rbrack
=
\left\lbrack{\begin{array}{c}
  -\left\lbrack I_0^s\right\rbrack \\
  \left\lbrack I_0^s\right\rbrack + \left\lbrack I_{sp}\right\rbrack \\
  \end{array} } \right\rbrack. 
\label{eq:KCL_matrices}
\end{equation}
Using the superposition principle in circuit theory, i.e., the contributions due to different voltage/current sources can be considered separately, we can determine the contributions due to SMR and spin pumping accordingly.

Figure~\ref{fig:ST-FMR_circuit}(b) shows the charge circuit counterpart comprising the spin current dependent charge current sources due to ISHE and the conductance $G_0$ of the SHE layer. A conductance in case the magnet is metallic can be included by putting the conductance in parallel to $G_0$ to account for the current shunting.~\cite{roy17_2} The charge current sources in the Fig.~\ref{fig:ST-FMR_circuit}(b) can be expressed as 
\begin{equation}
I_0^c = \beta G_0 \left(V_3^z - V_4^z \right).
\label{eq:I_0c}
\end{equation}
Applying KCL at node 1 of the charge-circuit in Fig.~\ref{fig:ST-FMR_circuit}(b), we get
\begin{equation}
I_1^c = -I_0^c + G_0 \left(V_1^c - V_2^c \right).
\label{eq:I_1c}
\end{equation}

The contribution due to SMR is given by the spin-circuit in the Fig.~\ref{fig:ST-FMR_circuit}(c) as follows.
\begin{equation}
 \left\lbrack {\begin{array}{cc}
   \left\lbrack G_1\right\rbrack + \left\lbrack G_2\right\rbrack & -\left\lbrack G_2\right\rbrack \\
   -\left\lbrack G_2\right\rbrack & \left\lbrack G_1\right\rbrack + \left\lbrack G_2\right\rbrack + \left\lbrack G^{\uparrow\downarrow}\right\rbrack
 \\
  \end{array} } \right\rbrack
\left\lbrack{\begin{array}{c}
  \left\lbrack V_3^s\right\rbrack \\
  \left\lbrack V_4^s\right\rbrack \\
  \end{array} } \right\rbrack
=
\left\lbrack{\begin{array}{c}
  -\left\lbrack I_0^s\right\rbrack \\
  \left\lbrack I_0^s\right\rbrack\\
  \end{array} } \right\rbrack.
\label{eq:KCL_SMR}
\end{equation}

After solving the above equations in the matrix form, we get
\begin{equation}
V_3^m - V_4^m = - \frac{2\,\mathrm{tanh}(t/2\lambda)}{G_\lambda} I_{0}^{s0} m_z,
\label{eq:comp_m}
\end{equation}
\begin{equation}
V_3^{m'} - V_4^{m'} = \frac{1}{G_\lambda} Im\lbrace F(t)\rbrace I_{0}^{s0} \frac{1}{\omega},
\label{eq:comp_m'}
\end{equation}
and
\begin{align}
&V_3^{m\times m'} - V_4^{m\times m'} = - \frac{\mathrm{tanh}(t/\lambda)}{G_\lambda} I_{0}^{s0} \frac{1}{\omega} \nonumber\\ 
&-\frac{(G_\lambda \mathrm{tanh}(t/\lambda) +2G_r)\,\mathrm{tanh}^2(t/\lambda)\,\mathrm{tanh}^2(t/2\lambda)}{(G_\lambda \mathrm{tanh}(t/\lambda) +2G_r)^2 + 4 G_i^2} I_{0}^{s0} \frac{1}{\omega}
\label{eq:comp_m_m'}
\end{align}
where 
\begin{equation}
F(t)=\frac{(2G^{\uparrow\downarrow}/G_\lambda)\,\mathrm{tanh}^2(t/2\lambda)}{1+(2G^{\uparrow\downarrow}/G_\lambda)\,\mathrm{coth}(t/\lambda)}. 
\end{equation}

We can write the spin voltage differences in the longitudinal and transverse directions as follows (see supplementary material).
\begin{equation}
V_3^z - V_4^z = (V_3^m - V_4^m) m_z + (V_3^{m\times m'} - V_4^{m\times m'})\left(1-m_z^2\right) \omega,
\label{eq:comp_z}
\end{equation}
\begin{align}
V_3^x - V_4^x =& (V_3^m - V_4^m) m_x - (V_3^{m'} - V_4^{m'})\, m_y \,\omega \nonumber\\
							&- (V_3^{m\times m'} - V_4^{m\times m'})\,m_z m_x \,\omega.
\label{eq:comp_x}
\end{align}
From Equations~\eqref{eq:I_1c} and~\eqref{eq:comp_z}, and using the equality $ 2 \mathrm{tanh}(t/2\lambda) = \mathrm{tanh}(t/\lambda) \, (1+\mathrm{tanh}^2(t/2\lambda))$, we get
\begin{align}
\frac{I_1^c}{tw} = \frac{\sigma}{l}\left(V_1^c - V_2^c\right) &\left\lbrace 1 + \theta_{SH}^2 (\lambda/t) \left\lbrack 2\mathrm{tanh}(t/2\lambda)\right.\right. \nonumber\\
& - \left.\left. (1-m_z^2)\,Re\lbrace F(t)\rbrace \right\rbrack\right\rbrace.
\label{eq:SMR_long}
\end{align}
Using $I_5^c = \beta G_0 \left(V_3^x - V_4^x \right)$ and Equation~\eqref{eq:comp_x}, we get
\begin{equation}
\frac{I_5^c}{lt} = -\theta_{SH}^2 \frac{\sigma\lambda}{lt}\left(V_1^c - V_2^c\right) \left\lbrack m_z m_x \,Re + m_y\, Im \right\rbrack \, F(t).
\label{eq:SMR_tran}
\end{equation}
The Equations~\eqref{eq:SMR_long} and~\eqref{eq:SMR_tran} are known as \emph{longitudinal} and \emph{transverse} SMR, respectively and these results match the existing results in the literature.~\cite{RefWorks:1024,RefWorks:1253} Note that the SMR is proportional to the square of the spin Hall angle ($\theta_{sh}^2$) due to the coordinated action of SHE and ISHE, and therefore materials with higher values of $\theta_{sh}$ would aid in increasing the SMR.

\begin{figure}
\centering
\includegraphics[width=0.45\textwidth]{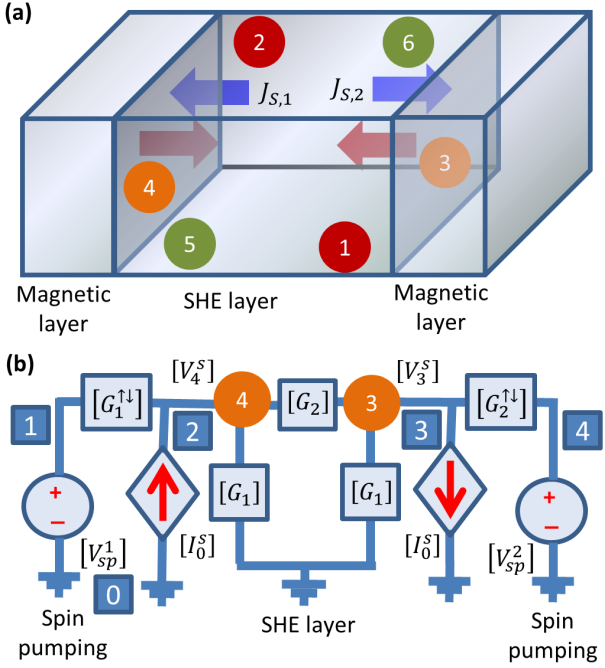}
\caption{\label{fig:SMR_spin_valve} (a) A spin-valve structure employing SMR i.e., a trilayer with an SHE layer sandwiched by two magnetic insulator layers. The SMR will be dependent on the magnetization directions of the two magnetic layers. (b) The corresponding spin-circuit representation employing the two interfacial spin-mixing conductances $G_1^{\uparrow\downarrow}$ and $G_2^{\uparrow\downarrow}$. and two spin batteries $[V_{sp}^1]$ and $[V_{sp}^2]$. The spin-circuit representation of the SHE layer remains same as shown in the Fig.~\ref{fig:ST-FMR_circuit}(a). Charge potentials are developed at the surfaces marked by 1 and 2, and marked by 5 and 6 and the charge-circuit as presented in the Fig.~\ref{fig:ST-FMR_circuit}(b) remains same.}
\end{figure}

The contribution due to spin pumping is given by the spin-circuit in the Fig.~\ref{fig:ST-FMR_circuit}(d) as follows.
\begin{equation}
 \left\lbrack {\begin{array}{cc}
   \left\lbrack G_1\right\rbrack + \left\lbrack G_2\right\rbrack & -\left\lbrack G_2\right\rbrack \\
   -\left\lbrack G_2\right\rbrack & \left\lbrack G_1\right\rbrack + \left\lbrack G_2\right\rbrack + \left\lbrack G^{\uparrow\downarrow}\right\rbrack
 \\
  \end{array} } \right\rbrack
\left\lbrack{\begin{array}{c}
  \left\lbrack V_3^s\right\rbrack \\
  \left\lbrack V_4^s\right\rbrack \\
  \end{array} } \right\rbrack
=
\left\lbrack{\begin{array}{c}
  0 \\
  \left\lbrack I_{sp}\right\rbrack \\
  \end{array} } \right\rbrack.
\label{eq:KCL_spin_pumping}
\end{equation}
After solving the above equations in the matrix form, we get
\begin{equation}
\left . V_3^{m'} - V_4^{m'} \right|_{sp} = (\hbar/2e) Im\lbrace \eta(t)\rbrace,
\label{eq:comp_m'_sp}
\end{equation}
and
\begin{equation}
\left . V_3^{m\times m'} - V_4^{m\times m'} \right|_{sp} = (\hbar/2e) Re\lbrace \eta(t)\rbrace
\label{eq:comp_m_m'_sp}
\end{equation}
where 
\begin{equation}
\eta(t)=\frac{(2G^{\uparrow\downarrow}/G_\lambda)\,\mathrm{tanh}(t/2\lambda)}{1+(2G^{\uparrow\downarrow}/G_\lambda)\,\mathrm{coth}(t/\lambda)}.
\end{equation}
Note that there is no contribution in the direction of $\mathbf{m}$ since we are not considering any spin-wave propagation or magnons. From Equation~\eqref{eq:I_0c}, we get
\begin{equation}
\left . \frac{I_0^c}{tw} \right|_{sp} = \frac{\hbar}{2e} \theta_{SH} \frac{\sigma }{t} \left\lbrack Re\lbrace \eta(t)\rbrace \left(\mathbf{m} \times \mathbf{m'}\right)_z + Im\lbrace \eta(t)\rbrace \mathbf{m'}_z \right\rbrack,
\label{eq:spin_pumping}
\end{equation}
which matches the mathematical expression derived in literature.~\cite{RefWorks:1025} Note that in the ST-FMR experiments, the longitudinal SMR and spin pumping contributions can be separated from the involved angular dependence.

For a more complex system e.g., a spin-valve employing SMR~\cite{RefWorks:1024} as shown in the Fig.~\ref{fig:SMR_spin_valve}, the analytical expressions become tedious and this clearly signifies the prowess of the spin-circuit approach. We can simply write a netlist [see the node numbers in squares in the Fig.~\ref{fig:SMR_spin_valve}(b)] as follows and solve that using a circuit solver.
\begin{lstlisting}[mathescape,columns=fullflexible,basicstyle=\fontfamily{lmvtt}\selectfont,]
	conductances = [1 2 $G_1^{\uparrow\downarrow}$; 3 4 $G_2^{\uparrow\downarrow}$;
			2 0 $[G_1]$; 2 3 $[G_2]$; 3 0 $[G_1]$];
	voltageSources = [1 0 $[V_{sp}^1]$; 4 0 $[V_{sp}^2]$];
	currentSources = [0 2 $[I_0^s]$; 3 0 $[I_0^s]$];
\end{lstlisting}
While using a circuit solver, there is also the other way that we can follow i.e., convert $G_1^{\uparrow\downarrow}$ and $G_2^{\uparrow\downarrow}$ to $z$-$x$-$y$ basis, but mathematically working with the $z$-$x$-$y$ basis is tedious and therefore is not followed for the derivation of analytical expressions here. 

To summarize, we have developed the spin-circuit representation of ST-FMR and have shown that both the SMR and spin pumping contributions  match the established mathematical expressions in literature. Such circuits when complex can be simply solved using a circuit solver and therefore helps in analysis, benchmarking experimental results, and proposing creative designs of functional devices. We do need to consider the interfacial spin memory loss,~\cite{roy20_1,RefWorks:2602,RefWorks:2635,RefWorks:2634,RefWorks:1351,RefWorks:1114} which lowers the conductivity. Recent calculations~\cite{roy20_1,roy17_3,roy_spie_2018x,RefWorks:2589,RefWorks:2590,RefWorks:2609} signify the Elliott-Yafet spin relaxation mechanism in heavy metals for which $\lambda$ is dependent on thickness as conductivity varies with thickness of the samples. ST-FMR can allow us to understand and estimate the relevant parameters in the system and such understandings can benefit the device design using SHE,~\cite{roy14_3,RefWorks:2588} which has potential for building future spintronic devices, alongwith other promising energy-efficient emerging devices.~\cite{roy16_spin}

\vspace*{2mm}
See the supplementary material for detailed derivation.

\vspace*{2mm}
This work was supported by Science and Engineering Research Board (SERB) of India via sanction order SRG/2019/002166.

\vspace*{1mm}
The data that support the findings of this study are available from the corresponding author upon reasonable request.

\end{document}


\title{Supplementary Information\\Spin-circuit representation of spin-torque ferromagnetic resonance}

\author{Kuntal Roy}
\email{kuntal@iiserb.ac.in}
\noaffiliation
\affiliation{Department of Electrical Engineering and Computer Science\\
Indian Institute of Science Education and Research Bhopal, Bhopal, Madhya Pradesh 462066, India}



\maketitle


Here we provide the detailed derivations of some of the equations in the main Letter.
\bigskip

\textit{Derivation of Equation~(3).} We need to convert $\left\lbrack I_{0,zxy}^s\right\rbrack$ in $z$-$x$-$y$ basis as in the Equation~(2) of the main Letter to $\left\lbrack I_{0}^s\right\rbrack$ in $(\mathbf{m},\mathbf{m'},\mathbf{m} \times \mathbf{m'})$ basis, where $'$ denotes the time derivative. We can write the following 
\begin{equation}
\mathbf{\hat{e}_z} = \cos\theta\,\mathbf{\hat{e}_r} - sin\theta\,\mathbf{\hat{e}_\theta}.
\end{equation}
Using $\cos\theta=m_z$, $\mathbf{\hat{e}_r}=\mathbf{m}$, $\mathbf{m'}=sin\theta\,\omega\,\mathbf{\hat{e}_\phi}$, $\mathbf{m} \times \mathbf{m'}=-sin\theta\,\omega\,\mathbf{\hat{e}_\theta}$, and $\omega = \phi'$, we get the Equation~(3) in the main Letter as follows.
\begin{equation}
\left\lbrack I_{0}^s\right\rbrack = I_{0}^{s0} m_z \mathbf{m} + I_{0}^{s0}\frac{1}{\omega}\mathbf{m} \times \mathbf{m'}.
\end{equation}

\bigskip

\textit{Derivation of Equation~(4).} In Fig.~2(a), we apply Kirchhoff's current law (KCL) at the nodes 3 and 4, and get the following equations, respectively.
\begin{align}
\left\lbrack G_2\right\rbrack \left(\left\lbrack V_3^s\right\rbrack - \left\lbrack V_4^s\right\rbrack\right) + \left\lbrack G_1\right\rbrack \left\lbrack V_3^s\right\rbrack + \left\lbrack I_0^s\right\rbrack &= 0.\\
\left\lbrack G^{\uparrow\downarrow}\right\rbrack \left(\left\lbrack V_4^s\right\rbrack - \left\lbrack V_{sp}\right\rbrack\right) + \left\lbrack G_1\right\rbrack \left\lbrack V_4^s\right\rbrack + \left\lbrack G_2\right\rbrack \left(\left\lbrack V_4^s\right\rbrack - \left\lbrack V_3^s\right\rbrack\right) - \left\lbrack I_0^s\right\rbrack &= 0.
\end{align}

After putting the above equations in matrix form and using $\left\lbrack I_{sp}\right\rbrack=\left\lbrack G^{\uparrow\downarrow}\right\rbrack \left\lbrack V_{sp}\right\rbrack$, we get the Equation~(4) in the main Letter as follows.

\begin{equation}
 \left\lbrack {\begin{array}{cc}
   \left\lbrack G_1\right\rbrack + \left\lbrack G_2\right\rbrack & -\left\lbrack G_2\right\rbrack \\
   -\left\lbrack G_2\right\rbrack & \left\lbrack G_1\right\rbrack + \left\lbrack G_2\right\rbrack + \left\lbrack G^{\uparrow\downarrow}\right\rbrack
 \\
  \end{array} } \right\rbrack
\left\lbrack{\begin{array}{c}
  \left\lbrack V_3^s\right\rbrack \\
  \left\lbrack V_4^s\right\rbrack \\
  \end{array} } \right\rbrack
=
\left\lbrack{\begin{array}{c}
  -\left\lbrack I_0^s\right\rbrack \\
  \left\lbrack I_0^s\right\rbrack + \left\lbrack I_{sp}\right\rbrack \\
  \end{array} } \right\rbrack. 
\label{eq:KCL_matrices}
\end{equation}

\bigskip

\textit{Derivation of Equations~(12) and~(13).} We can determine the $z$- and $x$-components of the spin potentials developed in $(\mathbf{m},\mathbf{m'},\mathbf{m} \times \mathbf{m'})$ basis as follows.
\begin{align}
V^z &= V^m cos\theta + V^{m\times m'} sin^2\theta\, \omega \nonumber\\
		&= V^m m_z + V^{m\times m'}\left(1-m_z^2\right) \omega.
\label{eq:comp_z}
\end{align}
\begin{align}
V^x &= V^m sin\theta\,cos\phi - V^{m'} sin\theta\,sin\phi \,\omega  - V^{m\times m'} cos\theta\,sin\theta\,cos\phi\,\omega \nonumber\\
		&= V^m m_x - V^{m'} m_y \,\omega - V^{m\times m'}\,m_z m_x \,\omega.
\label{eq:comp_x}
\end{align}

After taking the differences in the spin potentials developed at the nodes 3 and 4, we get the Equations~(12) and~(13) in the main Letter as follows.
\begin{equation}
V_3^z - V_4^z = (V_3^m - V_4^m) m_z + (V_3^{m\times m'} - V_4^{m\times m'})\left(1-m_z^2\right) \omega.
\label{eq:comp_z_diff}
\end{equation}
\begin{equation}
V_3^x - V_4^x = (V_3^m - V_4^m) m_x - (V_3^{m'} - V_4^{m'})\, m_y \,\omega - (V_3^{m\times m'} - V_4^{m\times m'})\,m_z m_x \,\omega.
\label{eq:comp_x_diff}
\end{equation}